\documentstyle[amssymb,aps,multicol,epsf]{revtex}

\begin{document}
\title{Spectra of complex networks}
\author{S. N. Dorogovtsev$^{1,3,\ast }$, A. V. Goltsev$^{2,3,\dagger }$, J. F. F.
Mendes$^{2,\ddagger}$ and A. N. Samukhin$^{2,3,\S}$}
\address{$^{1}$ Departamento de F\'{\i}sica and Centro de F\'{\i}sica do 
Porto, 
Faculdade de Ci\^{e}ncias, Universidade do Porto, \\
Rua do Campo Alegre 687, 4169-007 Porto, Portugal\\
$^{2}$ Departamento de F\'{\i}sica, Universidade 
de Aveiro, Campus Universit\'{a}rio de Santiago, 3810-193 Aveiro, Portugal\\
$^{3}$ A.F. Ioffe Physico-Technical Institute, 194021 St. Petersburg, Russia}
\maketitle

\begin{abstract}
We propose a general approach to the description of spectra of complex
networks. 
For the spectra of networks with uncorrelated vertices (and a local
tree-like structure), exact equations are derived. These equations are
generalized to the case of networks with correlations between neighboring
vertices. 
The tail of the density of eigenvalues $\rho \left( \lambda \right) $ at
large $\left| \lambda \right| $ is related to the behavior of the vertex
degree distribution $P\left( k\right) $ at large $k$. In particular, as $%
P\left( k\right) \sim k^{-\gamma }$, $\rho \left( \lambda \right) \sim
\left| \lambda \right| ^{1-2\gamma }$. We propose a simple approximation,
which enables us to calculate spectra of various graphs analytically. We
analyse spectra of various complex networks and discuss the role of vertices
of low degree. 
We show that spectra of locally tree-like random graphs may serve as a
starting point in the analysis of spectral properties of real-world
networks, e.g., of the Internet.
\end{abstract}




\begin{multicols}{2}

\narrowtext


\section{Introduction}

Many real-world technological, social and biological complex systems have a
network structure. Due to their importance and influence on our life
(recall, e.g., the Internet, the WWW, and genetic networks) investigations
of properties of complex networks are attracting much attention \cite
{ba99,s01,ab01a,dm01c,n03,w99,dm03}. Such properties as robustness against
random damages and absence of the epidemic threshold in the so called
\textquotedblleft scale-free\textquotedblright\ networks are nontrivial
consequences of their topological structure. Despite undoubted advances in
uncovering the main important mechanisms, shaping the topology of complex
networks, we are still far from complete understanding of all peculiarities
of their topological structure. That is why it is so important to look for
new approaches which can help us to reveal this structure.

The structure of networks may be completely described by the associated
adjacency matrices. The adjacency matrices of undirected graphs are
symmetric matrices 
with matrix elements, equal to number of edges between the given vertices.
The eigenvalues of an adjacency matrix are related to many basic topological
invariants of networks such as, for example, the diameter of a network \cite
{crs97,chung}. Recently, in order to characterize networks, it was proposed
to study spectra of eigenvalues of the adjacency matrices as a fingerprint
of the networks \cite{fff99,sfff03,m99,fdbv01,gkk01,esms02,vhe02,g03}. The
rich information about the topological structure and diffusion processes can
be extracted from the spectral analysis of the networks. Studies of spectral
properties of the complex networks may also have a general theoretical
interest. The random matrix theory has been successfully \ used to model
statistical properties of complex classical and quantum systems such as
complex nucleus, disordered conductors, chaotic quantum systems (see, for
example, reviews \cite{gmw98}), the glassy relaxation \cite{bg88} and so on.

As the adjacency matrices are random, in the limit $N\longrightarrow \infty $
($N$ is the total number of vertices), the density of eigenvalues could be
expected to converge to the semicircular distribution in accordance with the
Wigner theorem \cite{wigner}. However, Rodgers and Bray have demonstrated
that the density of eigenvalues of a sparse random matrix deviates from the
Wigner semicircular distribution and has a tail at large eigenvalues \cite
{rb88}, see also \cite{sc02}. Recent numerical calculations of the spectral
properties of small-world and scale-free networks \cite{m99,fdbv01,gkk01},
and the spectral analyses\ of the Internet \cite{fff99,sfff03,esms02,vhe02}
have also revealed that the Wigner theorem does not hold. The spectra of the
Internet \cite{fff99,sfff03} and scale-free networks \cite{fdbv01,gkk01}
demonstrate an unusual power-law tail in the region of large eigenvalues. At
the present time there is a fundamental lack in understanding of these
anomalies. In order to carry out a complete spectral analysis of real
networks it is necessary to take into account all features of these complex
systems described by a degree distribution, degree correlations, the
statistics of loops, etc. At this time there is no regular approach that
allows one to handle this problem. Our paper fills this gap.

Our approach is valid for any network which has a {\em local tree-like
structure}. In particular, these are uncorrelated random graphs with a given
degree distribution \cite{mr95,dms03}, and their straightforward
generalizations \cite{nmejb02} allowing pair correlations of the nearest
neighbors. These graph ensembles have one common property: almost every
finite connected subgraph of the infinite graph is a tree. The tree is a
graph, which has no loops. A random Bethe lattice is an infinite random
tree-like graph. All vertices on a Bathe lattice are statistically
equivalent \cite{note}. These features (the absence of loops and the
statistical equivalence of vertices) are decisive for our approach. The
advantage of Bethe lattices is that they frequently allow analytical
solutions for a number of problems: random walks, spectral problems, etc.

Real-world networks, however, often contain numerous loops. In particular,
this is reflected in a strong ``clustering'', which means that the
(relative) number of loops of length $3$ do not vanish even in very large
networks. 
Nevertheless, we believe, that the study of graphs with a local tree-like
structure may serve as a starting point in the description of more complex
network architectures.

In the present paper we will derive exact equations which determine the
spectra of infinite random uncorrelated and correlated random tree-like
graphs. For this, we use a method of random walks. We propose a method of an
approximate solution of the equations. We shall show that the spectra of
adjacency matrices of random tree-like graphs have a tail at large
eigenvalues. In the case of a scale-free degree distribution, the density of
eigenvalues has a power-law behavior. We will compare spectra of random
tree-like graphs and spectra of real complex networks. The role of weakly
connected vertices will also be discussed.

\section{General approach}

Let $\hat{A}=\left( a_{vw}\right) $ be the $N\times N$ symmetric adjacency
matrix of an $N$-vertex Mayer's graph $G$, $a_{vw}^{2}=a_{vw}$, $a_{vv}=0$
(the Mayer graph has either $0$ or $1$ edges between any pair of vertices,
and has no ``tadpoles'', i.e., edges attached at a single vertex). Degree $%
k_{v}$ (the number of connections) of a vertex $v$ is defined as 
\begin{equation}
k_{v}=\sum_{w}a_{vw}\,.  \label{constrain}
\end{equation}
A random graph, which is, in fact, an ensemble of graphs, is characterized
by a degree distribution $P(k)$: 
\begin{equation}
P\left( k\right) =\left\langle \frac{1}{N}\sum_{v=1}^{N}\delta \left(
k_{v}-k\right) \right\rangle \,.  \label{1}
\end{equation}
Here, $\langle\ \rangle$ is the averaging over the ensemble. We suppose that
each graph in the ensemble has $N$ vertices. Graph ensembles with a given 
{\em uncorrelated} vertex degree distribution may be realized, e.g., as
follows. Consider all possible graphs with a sequence of the numbers $%
\{N\left(k\right)\}$ of vertices of degree $k$, $k=1,2,\dots $, $\sum_k
N(k)=N$, assuming $N\left( k\right) /N \to P\left(k\right) $ in the
thermodynamic limit [$N(k)\to \infty$, $N \to \infty$]. Suppose that all
these graphs are equiprobable. Then, simple statistical arguments lead to
the conclusion that almost all finite connected subgraphs of an infinite
graph do not contain loops.

This approach can be easily generalized to networks with 
correlations between nearest-neighbor vertices, characterized by the
two-vertex degree distribution: 
\begin{equation}
P_{2}\left( k,k^{\prime }\right) =\left\langle \frac{1}{2L}%
\sum_{v,w=1}^{N}a_{vw}\delta \left( k_{v}-k\right) \delta \left(
k_{w}-k^{\prime }\right) \right\rangle \,.  \label{1a}
\end{equation}
Here $L=\left( 1/2\right) \sum_{v,w}a_{vw}$ is the total number of edges. In
the case of an uncorrelated graph we have 
\begin{equation}
P_{2}\left( k,k^{\prime }\right) = \frac{kk^\prime}{{\langle k \rangle}^2}
P\left( k\right) P\left( k^{\prime }\right) \,,  \label{1b}
\end{equation}
where $\langle k \rangle = 2L/N$ is the mean degree of a vertex.

The spectrum of $\widehat{A}$ may be calculated by using the method of
random walks on a tree-like graph $G$ and generating functions \cite{r01}.\
We define a generating function 
\begin{equation}
R(z)=\frac{1}{N}\sum_{v=1}^{N}\sum_{n=0}^{\infty }\rho _{v}(n)z^{n}\,,
\label{2}
\end{equation}
where $\rho _{v}(n)$ is the number of walks of length $n$ from $v$ to $v$,
where $v$ is any vertex of $G$: 
\begin{equation}
\rho _{v}(n)=(\widehat{A}^{n})_{v,v}\,.  \label{3}
\end{equation}
In a tree-like graph the number of steps $n$ is an even number. In order to
return to $v$ we must go back along all of the edges we have gone.

Let $q_{v}(n)$ be the number of walks of length $n$ starting at $v$ and
ending at $v$ for the first time. 
We define 
\begin{equation}
Q_{v}(z)=\sum_{n=0}^{\infty }q_{v}(n)z^{n}\,.  \label{4a}
\end{equation}
One can prove that 
\begin{equation}
R(z)=\frac{1}{N}\sum_{v=1}^{N}\frac{1}{1-Q_{v}(z)}\,.  \label{5}
\end{equation}
Let $d(w,v)=m\geqslant 1$ be the distance from $w$ to $v$ and $%
t_{w,v}^{(m)}(n)$ be the the number of paths of length $n$ starting at $w$
and ending at $v$ for the first time. We define 
\begin{equation}
T_{wv}^{(m)}(z)=\sum_{n=0}^{\infty }t_{w,v}^{(m)}(n)z^{n}\,.  \label{6}
\end{equation}
One can prove 
\begin{eqnarray}
Q_{v}(z) &=&z\sum_{w}T_{wv}^{(1)}(z)\,,  \label{7} \\
T_{wv}^{(m)}(z)
&=&T_{wg_{1}}^{(1)}(z)T_{g_{1}g_{2}}^{(1)}(z)...T_{g_{m-1}v}^{(1)}(z)\,,
\label{8}
\end{eqnarray}
where $w\longrightarrow g_{1}\longrightarrow g_{2}\longrightarrow
...g_{m-1}\longrightarrow v$ is the shortest path from $w$ to $v$. There is
an important relationship: 
\begin{eqnarray}
T_{wv}^{(1)}(z) &=&z+z\sum_{g}T_{gv}^{(2)}(z)  \nonumber \\
&=&z+z\sum_{g}T_{gw}^{(1)}(z)T_{wv}^{(1)}(z)\,.  \label{9b}
\end{eqnarray}
In this sum the vertex $g$ is the nearest neighbor of $w$ and a second
neighbor of the vertex $v$. Solving the recurrence equation (\ref{9b}), we
can find $T_{wv}^{(1)}(z)$ and $Q_{v}(z)$.

We define $\widetilde{T}_{wv}^{(1)}(z)\equiv T_{wv}^{(1)}(z^{-1})$. Equation
(\ref{9b}) may be written in a form 
\begin{equation}
\widetilde{T}_{wv}^{(1)}(z)=\frac{1}{z-\sum_{g}\widetilde{T}_{gw}^{(1)}(z)}%
\,.  \label{31}
\end{equation}
We can find $Q_{v}(z)$, from which we get $R(z).$ Let us define $B(z)\equiv
z^{-1}R(z^{-1})$. Then the density of the eigenvalues $\lambda $ of a random
graph is determined as follows: 
\begin{equation}
\rho (\lambda )=-%
\mathop{\rm Im}%
\left\langle B(\lambda +i\varepsilon ))\right\rangle /\pi \,,  \label{10}
\end{equation}
where $\varepsilon $ is positive and tends to zero. Note that the equations (%
\ref{4a})--(\ref{31}) are valid for both uncorrelated and correlated
tree-like graphs.

In the case of a $k-$regular\ connected graph we have $\widetilde{T}%
_{wv}^{(1)}(z)\equiv T(z)$ and $Q_{v}(z)\equiv Q(z)$. Eq. (\ref{31}) gives 
\begin{equation}
zT(z)-(k-1)T^{2}(z)=1\,.  \label{19a}
\end{equation}
Solving this equation, we get the well known result: 
\begin{equation}
\rho (\lambda )=\frac{k}{2\pi }\frac{\sqrt{4(k-1)-\lambda ^{2}} }{%
k^{2}-\lambda ^{2}} \,.  \label{20}
\end{equation}
This is a continuous spectrum of extended eigenstates with eigenvalues $%
\left\vert \lambda \right\vert <2\sqrt{k-1}$. The presence of the
denominator on the right-hand side of Eq. (\ref{20}) leads to a difference
of the spectrum of this graph from Wigner's semi-circular law. In exact
terms, Wigner's law is valid for the eigenvalue spectra of real symmetric
random matrices whose elements are independent identically distributed
Gaussian variables \cite{wigner}. These specific random matrices for
Wigner's law essentially differ from the adjacency matrices, which we
consider in this paper. So, in our case, the semicircular law may be used
only as a landmark for a contrasting comparison. 

\section{\protect\smallskip Spectra of uncorrelated graphs}

In the case of uncorrelated random tree-like graphs, $\ k_{w}-1$ random
parameters $\widetilde{T}_{gw}^{(1)}(z)$ on the right-hand side of Eq. (\ref
{31}) are equivalent and statistically independent. They are also
independent on the degree $k_{w}$. We define the distribution function of $%
\widetilde{T}_{wv}^{(1)}(z)$ at $z=\lambda +i\varepsilon $ in the Fourier
representation as: 
\begin{equation}
F_{\lambda }(x)=\left\langle \exp \left[ -ix\widetilde{T}_{wv}^{(1)}(\lambda
+i\varepsilon )\right] \right\rangle \,,  \label{32}
\end{equation}
where the brackets $\left\langle ...\right\rangle $ means the averaging over
the ensemble of random uncorrelated graphs associated with a degree
distribution $P(k)$.\ The statistical independence of the $k-1$ random
parameters $\widetilde{T}_{gw}^{(1)}(\lambda +i\varepsilon )\equiv T_{i}$, $%
i=1,2,...k-1$, $k\equiv k_{w}$, on the right hand side of Eq. (\ref{31})
allows us to use the following identity: 
\begin{eqnarray}
&&F_{\lambda }(x)\equiv \left\langle \exp (-ixT)\right\rangle =\left\langle
\exp (-\frac{ix}{\lambda +i\varepsilon -\sum_{i=1}^{k-1}T_{i}})\right\rangle 
\nonumber \\[5pt]
&=&1-\sqrt{x}\int\limits_{0}^{\infty }\frac{dy}{\sqrt{y}}J_{1}(2\sqrt{xy}%
)\left\langle \exp (iy[\lambda +i\varepsilon
-\sum_{i=1}^{k-1}T_{i}])\right\rangle   \nonumber \\[5pt]
&=&1-\sqrt{x}\int\limits_{0}^{\infty }\frac{dy}{\sqrt{y}}J_{1}(2\sqrt{xy}%
)e^{iy(\lambda +i\varepsilon )}\times   \nonumber \\[5pt]
&&\sum_{k}\frac{kP(k)}{\langle k\rangle }\left\langle \exp
(-iyT)\right\rangle ^{k-1}\,,  \label{identity}
\end{eqnarray}
where $J_{1}(x)$ is the Bessel function and $\left\langle k\right\rangle
=\sum_{k}kP(k)$. Thus, we get the exact self-consistent equation for $%
F_{\lambda }(x)$: 
\begin{equation}
F_{\lambda }(x)=1-\sqrt{x}\int_{0}^{\infty }\frac{dy}{\sqrt{y}}J_{1}(2\sqrt{%
xy})e^{iy\lambda }\Phi _{1}(F_{\lambda }(y))\,,  \label{34}
\end{equation}
where $\Phi _{1}(x)\equiv \sum_{k=1}^{\infty }kP(k)x^{k-1}/\left\langle
k\right\rangle $. Solving Eq. (\ref{34}) gives the distribution of $T$, and
\ so we can obtain $Q$, from which we get $R$. Eqs. (\ref{5}), (\ref{7}),
and (\ref{10}) give 
\begin{eqnarray}
\rho (\lambda ) &=&-\frac{1}{\pi }%
\mathop{\rm Im}%
\left\langle \frac{1}{\lambda -\sum_{i=1}^{k}T_{i}}\right\rangle   \nonumber
\\[5pt]
&=&\frac{1}{\pi }%
\mathop{\rm Re}%
\int_{0}^{\infty }dye^{iy\lambda }\Phi (F_{\lambda }(y))\,,  \label{ro}
\end{eqnarray}
where $\Phi (x)\equiv \sum_{k=1}^{\infty }P(k)x^{k}$. From Eq. (\ref{34}),
we find the $n$-th moment of the distribution function $\Psi _{\lambda }(T)$%
, Eq. (\ref{32}): 
\begin{equation}
M_{n}\equiv \left\langle T^{n}\right\rangle =\frac{1}{(n-1)!\,i^{n}}%
\int_{0}^{\infty }dy\,y^{n-1}e^{iy\lambda }\Phi _{1}(F_{\lambda }(y))\,.
\label{44}
\end{equation}

\section{Effective medium approximation}

In a general case it is difficult to solve Eq. (\ref{34}) exactly. Let us
find an approximate solution. We neglect fluctuations of $T$ around a mean
value $T(\lambda )\equiv \left\langle T\right\rangle $. A self-consistent
equation for the function $T(\lambda )$ may be obtained if we insert 
\begin{equation}
F_{\lambda }(x)\approx e^{-ixT(\lambda )}  \label{35}
\end{equation}
into the right-hand side of Eq. (\ref{44}) for $n=1$. We get 
\begin{equation}
T(\lambda )=\frac{1}{\left\langle k\right\rangle }\sum_{k}\frac{kP(k)}{%
\lambda +i\varepsilon -(k-1)T(\lambda )}\,.  \label{23}
\end{equation}
Below we will call this approach an ``effective medium'' (EM) approximation.
At real $\lambda $, $T(\lambda )$ is a complex function, which is to be
understood as an analytic continuation from the upper half-plane of $\lambda 
$, $T\left( \lambda \right) \equiv T\left( \lambda +i\varepsilon\right) $.
Therefore, $%
\mathop{\rm Im}%
T(\lambda +i\varepsilon )<0$. In the framework of the EM approach, the
density $\rho (\lambda )$, Eq. (\ref{ro}), takes an approximate form 
\begin{equation}
\rho (\lambda )=-\frac{1}{\pi }\sum_{k}\frac{kP(k)%
\mathop{\rm Im}%
T(\lambda )}{(\lambda -k%
\mathop{\rm Re}%
T(\lambda ))^{2}+k^{2}(%
\mathop{\rm Im}%
T(\lambda ))^{2}} \,.  \label{41}
\end{equation}

\section{Tail behavior and finite-size effect}

Equation (\ref{23}) may be solved analytically at $\left| \lambda \right|
\gg 1$. We look for a solution in the region $\mathop{\rm Im}T(\lambda )\ll %
\mathop{\rm Re}T(\lambda )\ll 1$. It is convenient to use a continuum
approximation in Eq. (\ref{23}). The real and imaginary parts of this
equation take a form 
\begin{eqnarray}
&&\mathop{\rm Re}T(\lambda )=\frac{1}{2\lambda \left\langle k\right\rangle }%
\times   \nonumber \\[0.07in]
&&\int_{k_{0}}^{k_{cut}}\!\!\!\!\!\!\!\!\frac{dk\,kP(k)}{\displaystyle\left(
\!1-(k-1)\mathop{\rm Re}\frac{T(\lambda )}{\lambda }\right) ^{\!2}\!+\left(
\!(k-1)\mathop{\rm Im}\frac{T(\lambda )}{\lambda }\right) ^{\!2}}\,,
\label{1new} \\[0.1in]
&&1=\frac{1}{\lambda ^{2}\left\langle k\right\rangle }\times   \nonumber \\%
[5pt]
&&\int_{k_{0}}^{k_{cut}}\!\!\!\!\!\!\!\!\frac{dk\,k(k-1)P(k)}{\displaystyle%
\left( \!1-(k-1)\mathop{\rm Re}\frac{T(\lambda )}{\lambda }\right)
^{\!2}\!+\left( \!(k-1)\mathop{\rm Im}\frac{T(\lambda )}{\lambda }\right)
^{\!2}}\,,  \label{2new}
\end{eqnarray}
where $k_{0}$ and $k_{cut}$ are the smallest and largest degrees,
respectively. A region $k_{0}\leq k\ll k_{\lambda }$ gives a regular
contribution into the integrals (\ref{1new}) and (\ref{2new}) while a region 
$k\sim k_{\lambda }\gg 1$ gives a singular contribution. Here $k_{\lambda
}\equiv \lambda /\mathop{\rm Re}T(\lambda )+1$. As a result we obtain 
\begin{eqnarray}
\mathop{\rm Re} &&T(\lambda )\cong \frac{1}{2\lambda }+\frac{\pi k_{\lambda
}P(k_{\lambda })}{2\left\langle k\right\rangle \mathop{\rm Im}T(\lambda )}\,,
\label{3new} \\[7pt]
1\cong \frac{1}{\lambda ^{2}\left\langle k\right\rangle } &&%
\int_{k_{0}}^{k_{\lambda }}dk\,k(k-1)P(k)+\frac{\pi \lambda k_{\lambda
}P(k_{\lambda })}{\left\langle k\right\rangle \mathop{\rm Im}T(\lambda )}\,.
\label{4new}
\end{eqnarray}
If $P(k)$ decreases faster than $k^{-2}$ at $k\gg 1$, i.e. $\left\langle
k\right\rangle $ is finite, then in the leading order of $1/\lambda $ we
find 
\begin{equation}
T(\lambda )\cong \lambda ^{-1}-i\pi \left| \lambda \right| k_{\lambda
}P(k_{\lambda })/\left\langle k\right\rangle \,.  \label{39a}
\end{equation}
Within the same approach one can find from Eq. (\ref{41}) that the density $%
\rho (\lambda )$ also has two additive contributions 
\begin{equation}
\rho (\lambda )\cong -\frac{\left\langle k\right\rangle 
\mathop{\rm Im}%
T(\lambda )}{\pi \lambda ^{2}}+\frac{k_{\lambda }P(k_{\lambda })}{\left|
\lambda \right| }\,.  \label{5new}
\end{equation}
Inserting Eq. (\ref{39a}) gives the density 
\begin{equation}
\rho (\lambda )\cong 2\,\frac{k_{\lambda }}{|\lambda |}P(k_{\lambda })\,.
\label{39b}
\end{equation}
Here $k_{\lambda }=\lambda /\mathop{\rm Re}T(\lambda )+1=\lambda ^{2}+O(1)$.


The asymptotic expression (\ref{39b}) is our main result. 
The right-hand side of this expression originates from two equal, additive
contributions: the contribution from the real part of $T(\lambda)$ and the
one from the imaginary part of $T(\lambda)$. One can show that the
asymptotic behavior of the real part, $\mathop{\rm Re}T(\lambda )=\lambda
^{-1}+O(\lambda ^{-3})$, in the leading order of $1/\lambda $ is universal
and is valid even for graphs with finite loops. Contrastingly, the
asymptotics of 
$\mathop{\rm Im}T(\lambda )$ in the leading order of $1/\lambda $ and the
corresponding contribution to the right-hand side of Eq. (\ref{39b}) depend
on details of the structure of a network.


The analysis of Eq. (\ref{23}) shows that the main contribution to an
eigenstate with a large eigenvalue $\lambda $ is given by vertices with a
large degree $k \sim k_{\lambda }\gg 1$. As we shall show below, in the
limit $\lambda \gg 1$, the result (\ref{39b}) is asymptotically exact. 
The relationship between largest eigenvalues and highest degrees, $\lambda^2
+ O(1) = k$, for a wide class of graphs was obtained in a 
mathematical paper, Ref. \cite{mp02}. 
This contribution of highly connected vertices may be compared with a simple
spectrum of ``stars'', which are graphs consisted of a vertex of a degree $k$%
, connected to $k$ dead ends. The spectrum consists of two eigenvalues $%
\lambda =\pm \sqrt{k}$ and a $(k-1)$-degenerate zero eigenvalue. Note that
asymptotically, in the limit of large $\lambda $, Eq. (\ref{39b}) gives $%
\rho (\lambda )\cong 2|\lambda |P(\lambda ^{2})$ if $P(k)$ decreases slower
than an exponent function at large $k$, that is, if higher moments of the
degree distribution diverge.

A classical random graph \cite{sr51,er} has the Poisson degree distribution $%
P(k)=e^{-\left\langle k\right\rangle }\left\langle k\right\rangle ^{k}/k!$.
The tail of $\rho (\lambda )$ is given by Eq. (\ref{39b}) with $k_{\lambda }
= \lambda ^{2}+a\gg 1$ where $a$ is a number of the order of $1$: 
\begin{equation}
\rho (\lambda )\sim \lambda ^{-2(\lambda ^{2}+a)}\exp [(1+\ln \langle
k\rangle )\lambda ^{2}].  \label{tailP}
\end{equation}
This equation agrees with the previous results \cite{rb88,sc02} obtained by
different analytical methods.

For a ``scale-free''\ graph with $P(k)\approx P_{0}k^{-\gamma }$ at large $k$%
, at $\left| \lambda \right| \gg 1$, we get an asymptotically exact
power-law behavior: 
\begin{equation}
\rho (\lambda )\cong 2\left| \lambda \right| P(\lambda ^{2})=2P_{0}\left|
\lambda \right| ^{-\delta }\,,  \label{42}
\end{equation}
where the eigenvalue exponent $\delta =2\gamma -1$.

At a finite $N\gg 1$, there is a finite-size cutoff of the degree
distribution $k_{cut}\sim k_{0}N^{1/(\gamma -1)}$ \cite{bk02}. The cutoff
determines the upper boundary of eigenvalues: $\lambda <k_{cut}^{1/2}$. This
result agrees with an estimation of the largest eigenvalue of sparse random
graphs obtained in Ref. \cite{ks03}.

Let us analyze the accuracy of the EM approach. One can use the following
criterion. We introduce a quantity $q_{n}\equiv M_{n}/T^{n}(\lambda )$.
Here, $T^{n}(\lambda )$ is the $n$-th moment of the approximate distribution
(\ref{35}). Inserting the function (\ref{35}) into Eq. (\ref{44}) gives $%
M_{n}$. The function $F_{\lambda }(x)=e^{-ixT(\lambda )}$ would be an exact
solution of Eq. (\ref{34}) if $q_{n}=1$ for all $n\geqslant 1$. Note that at 
$n=1$ we have $q_{1}=1$, because this equality is the basic equation in the
framework of the EM approximation. At $\lambda \gg 1$ and $%
P(k)=P_{0}k^{-\gamma }$, in the leading order of $1/\lambda $, Eq. (\ref{44}%
) gives 
\begin{eqnarray}
q_{n} &\cong &1-Ank_{0}\lambda ^{-2}\ \ \ \ \ \ \ \ \ \ \ \,\text{ at }%
\gamma >3\,,  \label{48a} \\
[5pt] &\cong &1-Ank_{0}\lambda ^{-2}\ln \lambda \ \ \ \ \ \ \text{ at }%
\gamma =3\,,  \label{48b} \\
[5pt] &\cong &1-Ank_{0}^{\gamma -2}\lambda ^{-2(\gamma -2)}\ \ \text{ at }%
2<\gamma <3 \,,  \label{48c}
\end{eqnarray}
where $k_{0}$ is the smallest degree in $P(k)$ and $A$ is a numerical
factor. This estimation allows us to conclude that at $\lambda \gg
k_{0}^{1/2}$ the EM solution becomes asymptotically exact.

At small $\lambda \lesssim k_{0}^{1/2}$, the EM approximation are less
accurate. For example, at $\lambda =0$ for a scale-free network, we obtain $%
q_{n}=S_{n}/S_{1}^{n}$, where $S_{n}=\left\langle k(k-1)^{-n}\right\rangle
/\left\langle k\right\rangle $. Only at large $\gamma \gg 1$ and $n<\gamma $%
, the parameter $q_{n}$ is close to 1, i.e. $q_{n}=1+O(n/\gamma )$.

One can conclude that the EM approach gives a reliable result close to the
exact one in the range 
\begin{equation}
k_{0}^{1/2}\ll \lambda \ll k_{cut}^{1/2}=k_{0}^{1/2}N^{1/2(\gamma -1)}\,.
\label{49}
\end{equation}
\qquad \qquad

In our derivations we assumed the tree-like local structure of a network,
that is, the absence of finite-size loops in an infinite network. Loosely
speaking, this assumption may fail if the second moment $\langle
k^{2}\rangle $ of the degree distribution diverges. This can be seen from
the following simple arguments. The length of a typical loop is of the order
of the average shortest-path length of a network. Since the mean number of
the second-nearest neighbors in the infinite uncorrelated net is $\langle
k(k-1) \rangle$ and diverges if $\langle k^{2}\rangle$ diverges, the average
shortest-path length and the length of a typical loop are small and may turn
out to be finite even in the limit of an infinite net if $\langle
k^{2}\rangle$ diverges. In this situation, the result (whether there are
loops of finite length in the infinite network or not) is determined by the
size-dependence of the cut-off of the degree distribution \cite{dms03}. In
its turn, this dependence is determined by the specifics of an ensemble and
varies from network to network.

\section{\protect\smallskip Spectra of correlated graphs}

Many real-world networks are characterized by strong correlations between
degrees of vertices \cite{pvv01,ms02,rsmob02}. The simplest ones are
correlations between degrees of neighboring vertices. Let us study the
effect of degree correlations on spectra of random tree-like graphs.

Using the pair degree distribution (\ref{1a}), it is convenient to introduce
the conditional distribution that a vertex of degree $k$ is connected to a
vertex of degree $k_{1}$: 
\begin{equation}
P(k_{1}\left| k\right) =\left\langle k\right\rangle P(k_{1},k)/kP(k)\,.
\label{18}
\end{equation}
The method used above for the calculation of spectra of uncorrelated graphs
may be generalized to correlated graphs. For this, one should take into
account correlations between the degree of a vertex $v$ and the generating
function $\widetilde{T}_{wv}^{(1)}(z)$ in Eq. (\ref{31}). We define the
distribution function of $\widetilde{T}_{wv}$ in the Fourier representation
as: 
\begin{equation}
F_{\lambda }(k,x)=\left\langle \delta \left( k_{v}-k\right) \exp \left[ -ix%
\widetilde{T}_{wv}^{(1)}(\lambda +i\varepsilon )\right] \right\rangle .
\label{51}
\end{equation}
Averaging Eq. (\ref{31}) and using the identity (\ref{identity}), we obtain
an exact equation for $F_{\lambda }(k,x)$: 
\begin{eqnarray}
F_{\lambda }(k,x) &=&1-\sqrt{x}\sum_{k_{1}}P\left( k_{1}\right| k)  \nonumber
\\
&&\times \int_{0}^{\infty }\frac{dy}{\sqrt{y}}J_{1}(2\sqrt{xy})e^{iy\lambda
}F_{\lambda }^{k_{1}-1}(k_{1},y) \,,  \label{53}
\end{eqnarray}
The density of eigenvalues is of the following form 
\begin{equation}
\rho (\lambda )=\frac{1}{\pi }%
\mathop{\rm Re}%
\sum_{k}P(k)\int_{0}^{\infty }dy\, e^{iy\lambda }F_{\lambda }^{k}(k,y) \,.
\label{55}
\end{equation}
These equations are a generalization of the equations derived above for
uncorrelated graphs. Indeed, for an uncorrelated graph, we have $%
P(k_{1}\left| k\right) =kP(k)/\left\langle k\right\rangle $ and $F_{\lambda
}(k,x)=F_{\lambda }(x)$. As a result we get Eqs. (\ref{34}) and (\ref{ro}).

Let us use the EM approximation. We neglect fluctuations around a mean value 
$T(k,\lambda )$ and use an approximation 
\begin{equation}
F_{\lambda }(k,x)\approx e^{-ixT(k,\lambda )}.  \label{56}
\end{equation}
Then we get a self-consistent equation for the complex function $T(k,\lambda
)$: 
\begin{equation}
T(k,\lambda )=\sum_{k_{1}}\frac{P\left( k_{1}\right| k)}{\lambda
+i\varepsilon -(k_{1}-1)T(k_{1},\lambda )}\,.  \label{57}
\end{equation}
At $\lambda \gg 1$ this equation has a solution 
\begin{equation}
T(k,\lambda )=\lambda ^{-1}-i\pi \lambda P\left( k_{\lambda }\right| k)\,.
\label{58}
\end{equation}
This solution gives $\rho (\lambda )=2k_{\lambda }P(k_{\lambda })\,/\left|
\lambda \right| $ where $k_{\lambda } = \lambda ^{2}+O(1)$ as before. It
agrees with the result presented in Eq. (\ref{39b}) for uncorrelated graphs.
One concludes that the short-range correlations between degrees of
neighboring vertices in the scale-free networks does not change the
eigenvalue exponent $\delta =2\gamma -1$.

\section{Spectrum of a transition matrix}

Let us consider random walks on a graph with the transition probability $%
1/k_{v}$ of moving from a vertex $v$ to any one of its neighbors. The
transition matrix $\widehat{P}$ then satisfies 
\begin{equation}
P_{w,v}=a_{w,v}/k_{v}\,.  \label{61}
\end{equation}
Clearly, for each vertex $v$ 
\begin{equation}
\sum_{w}P_{w,v}=1\,.  \label{63}
\end{equation}
$\widehat{P}$ is related with the Laplacian of the graph 
\begin{equation}
L_{v,w} = \left\{ 
\begin{array}{cc}
\ \ \ \ \ \ 1 \ \ \ \ \ \  & \ \ \ \text{ if }v=u \\[5pt] 
-a_{v,w}/\sqrt{k_{v}k_{w}} & \text{ \ \ \ otherwise}
\end{array}
\right. ,  \label{62}
\end{equation}
as follows 
\begin{equation}
\widehat{P}=\widehat{D}^{1/2}(1-\widehat{L})\widehat{D}^{-1/2} \, ,
\label{PL}
\end{equation}
%
where $D_{v,w}=\delta _{v,w}/k_{v}$. Therefore, if we know the density $\rho
(\lambda )$ of eigenvalues of $\widehat{P}$, we can find the density of
eigenvalues of the Laplacian: $\rho _{L}(\lambda )=\rho (1-\lambda )$.

We denote the eigenvalues of the matrix $\widehat{P}$ by $\lambda
_{1}\geqslant \lambda _{2}\geqslant ...\geqslant \lambda _{N}$. The
eigenfunction $f_{v}=k_{v}$ corresponds to the largest eigenvalue $\lambda
_{1}=1$.

In order to calculate the spectrum of $\widehat{P}$ we use the same method
of random walks described in the Section II. The probability of one step is
given by Eq. (\ref{61}). We define the generating function $Q_{v}(z)$ and $%
\widetilde{T}_{wv}^{(1)}(z^{-1})\equiv T_{wv}^{(1)}(z)$ and obtain an exact
equation which is similar to Eq. (\ref{31}): 
\begin{equation}
\widetilde{T}_{wv}^{(1)}(z)=\frac{1}{k_{w}z-\sum_{g}\widetilde{T}%
_{gw}^{(1)}(z)}\,,  \label{64a}
\end{equation}
where $g\sim w$ but $g\neq v$. At $z=\lambda +i\varepsilon $, we get exact
equations for the function $F_{\lambda }(x)=\left\langle \exp (-ix\widetilde{%
T}_{wv}^{(1)}(\lambda +i\varepsilon ))\right\rangle $ and the density of the
eigenvalues $\rho (\lambda )$: 
\begin{equation}
F_{\lambda }(x)=1-\sqrt{x}\int_{0}^{\infty }\frac{dy}{\sqrt{y}}J_{1}(2\sqrt{%
xy})e^{iy\lambda }\Phi _{1}(e^{i\lambda y}F_{\lambda }(y))\,,  \label{64}
\end{equation}
\begin{equation}
\rho (\lambda )=\frac{1}{\pi }%
\mathop{\rm Re}%
\sum_{k}P(k)k\int_{0}^{\infty }dy\,e^{ik\lambda y}F_{\lambda }^{k}(y)\,.
\label{65}
\end{equation}
The function $F_{\lambda }(x)=e^{-ix}$ is an exact solution of Eq. (\ref{64}%
). This solution corresponds to the eigenvalue $\lambda _{1}=1$ \ and gives
the delta-peak $\delta (\lambda -1)$ in the density $\rho (\lambda )$. The
second largest eigenvalue $\lambda _{2}$ is related to several important
graph invariants such as the diameter $D$ of the graph, see, for example, 
\cite{chung}: 
\begin{equation}
D(G)\leqslant \frac{\log (N-1)}{\log (1/\lambda _{2})}\,.  \label{diam}
\end{equation}
Here the {\it diameter} of a graph is the maximum distance between any two
vertices of a given graph.

In order to find the spectrum at $\lambda \leqslant \lambda _{2}$ we use the
EM approach. We assume $F_{\lambda }(x)\approx e^{-ixT(\lambda )}$ and get
an equation for a complex function $T(\lambda )$ : 
\begin{equation}
T(\lambda )=\frac{1}{\left\langle k\right\rangle }\sum_{k}\frac{kP(k)}{%
k\lambda +i\varepsilon -(k-1)T(\lambda )}\,.  \label{66}
\end{equation}
$\rho (\lambda )$ is given by 
\begin{equation}
\rho (\lambda )=-\frac{1}{\pi }%
\mathop{\rm Im}%
\frac{1}{\lambda -T(\lambda )}\,.  \label{67}
\end{equation}
For completeness, we present the spectrum of the transition matrix of a $k$%
-regular tree: 
\begin{equation}
\rho (\lambda )=\frac{k}{2\pi } \frac{\sqrt{4(k-1)/k^{2}-\lambda ^{2}}}{%
1-\lambda ^{2}}\,,  \label{68}
\end{equation}
which easily follows from Eqs. (\ref{66}) and (\ref{67}). The second
eigenvalue is equal to $\lambda _{2}=2\sqrt{k-1}/k$.

\section{Analysis of spectra}

Let us compare available spectra of classical random graphs and scale-free
networks \cite{fdbv01,gkk01}, empirical spectra of the Internet \cite
{fff99,sfff03,vhe02}, and spectra of random tree-like graphs.

\begin{figure}
\epsfxsize=82mm
\centerline{\epsffile{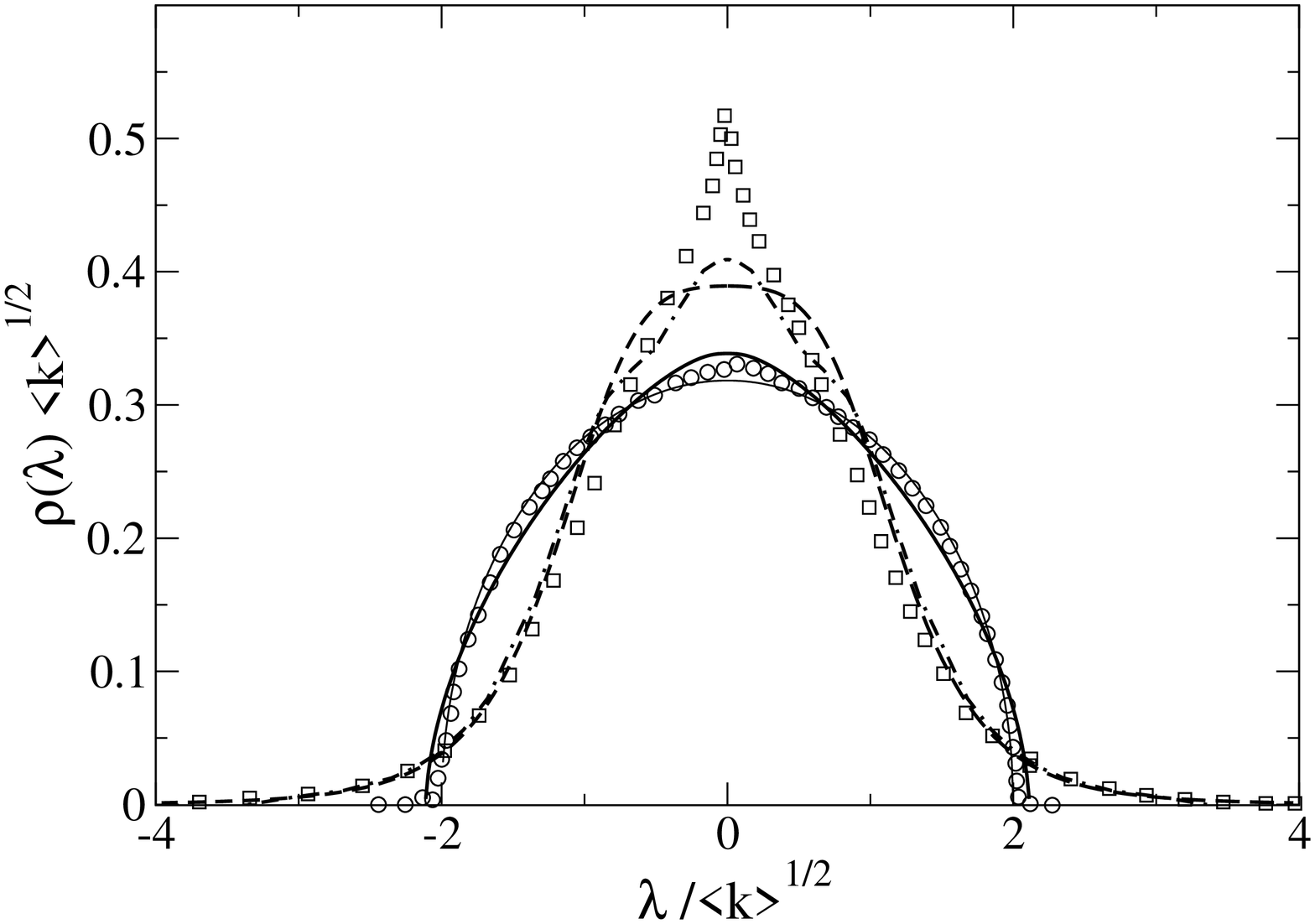}}
\caption{ Density of eigenvalues of the adjacency matrices of two networks.
(i) The classical random graph (the Erd\H{o}s-R\'{e}nyi model) with the
average degree $\left\langle k\right\rangle =10$: the effective medium (EM)
approach (the solid line) and numerical calculations for the graphs of
20\,000 vertices \protect\cite{fdbv01} (the open circles). (ii) The
scale-free random tree-like graph with $\protect\gamma =3$ and the smallest
degree $k_{0}=5$: the EM approach (the dashed line), the improved EM
approach, see the text (the dashed-dotted line). The results of the
simulations of the Barab\'{a}si-Albert model of 7000 vertices \protect\cite
{fdbv01} (the open squares). The semicircular law is shown by the thin solid
line. }
\label{fig1}
\end{figure}

\begin{figure}
\epsfxsize=75mm
\centerline{\epsffile{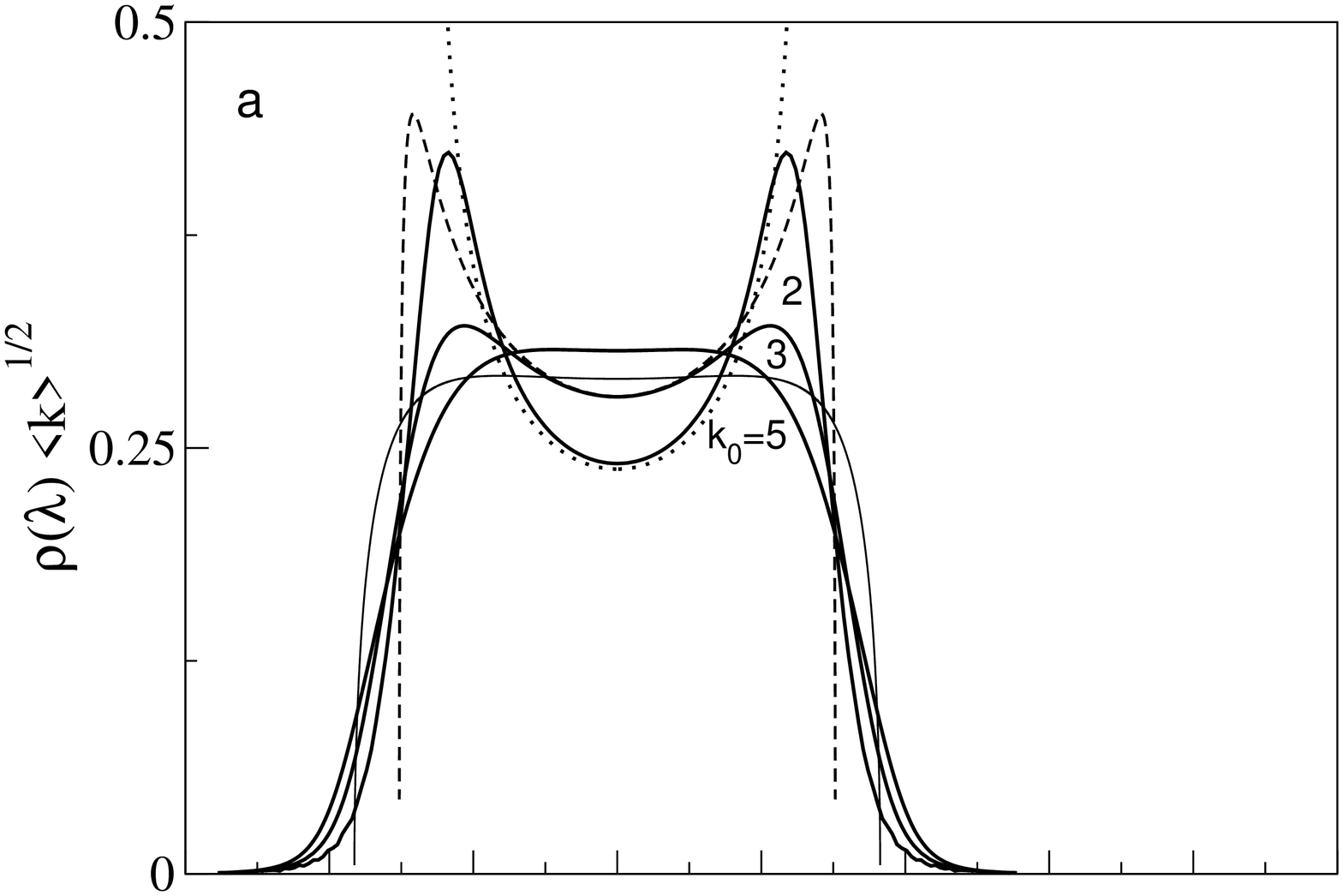}}
\epsfxsize=75mm
\centerline{\epsffile{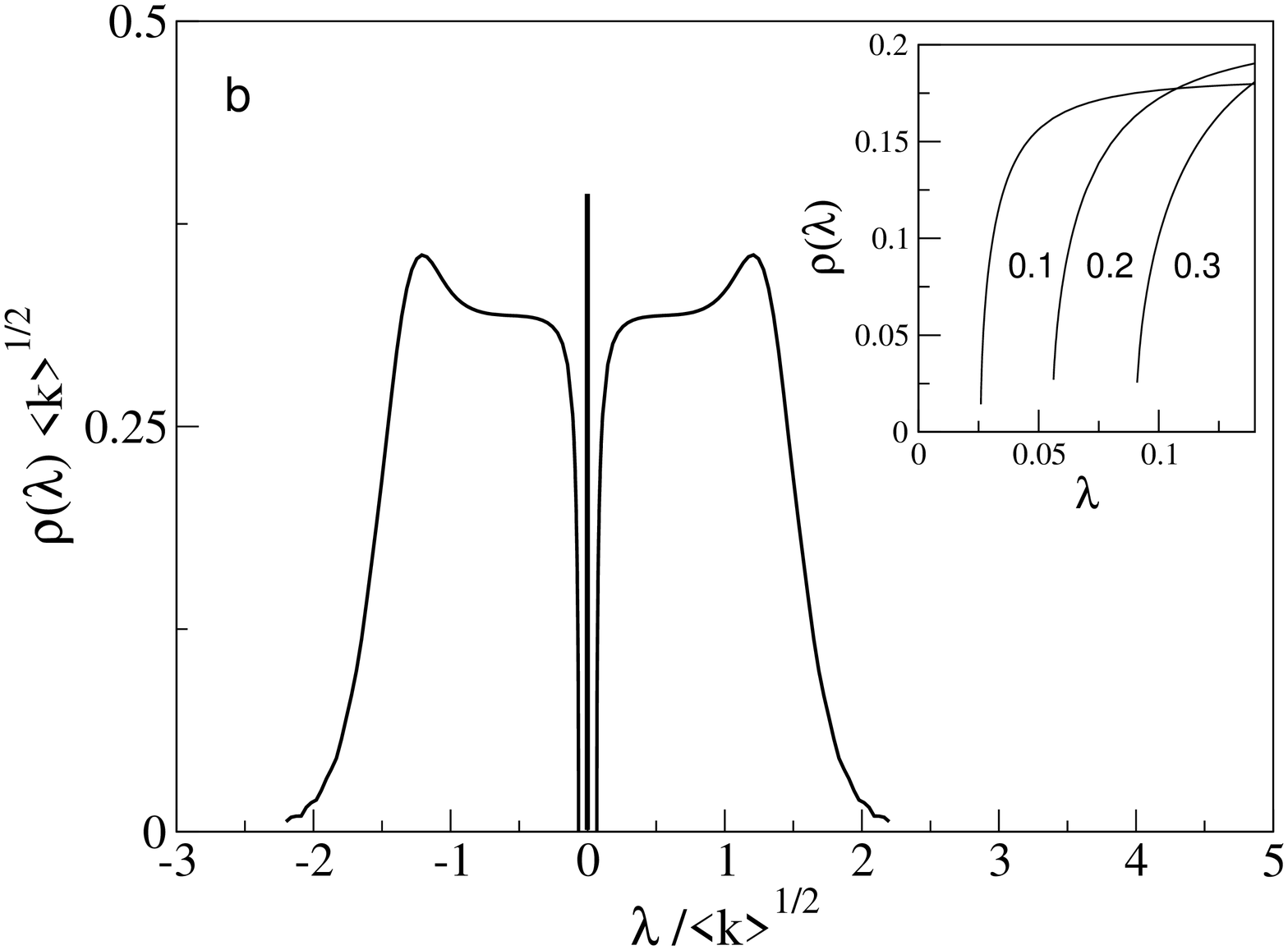}}
\caption{ Evolution of spectra of a random tree-like graph with the
scale-free degree distribution for $\protect\gamma =5$ and the smallest
degree $k_{0}=1,2,3$ and 5. The panel $a$ shows the spectra of the graphs
with $k_{0}=2,3$ and 5. The dotted line corresponds to the density of
eigenvalues of an infinite chain. The dashed and thin solid lines present
the spectrum of the $k=3$ and 6 regular Bethe lattices. The panel $b$ shows
the spectrum of a random uncorrelated graph having dead-end vertices with
the probability $P(1)=0.3$. The insert shows the behavior of \ the density
of eigenvalues $\protect\rho (\protect\lambda )$ near the dip at $%
P(1)=0.1,0.2$, and $0.3$. The central peak is produced by localized states. }
\label{fig2}
\end{figure}

At first we discuss spectra of adjacency matrices. The spectra were
calculated in the framework of the EM\ approach from Eqs. (\ref{23}) and (%
\ref{41}) for different degree distributions $P(k)$. Our results are
represented in Figs. \ref{fig1} and \ref{fig2}.

{\em Classical random graphs.} Classical random graphs have the Poisson
degree distribution. The density of eigenvalues of the associated adjacency
matrix has been obtained numerically in \cite{fdbv01}. In Fig. \ref{fig1} we
display results of the numerical calculations and our results obtained
within the EM approach. We found a good agreement in the whole range of
eigenvalues. There are only some small differences in the region of small
eigenvalues which may be explained by an inaccuracy of the EM approach in
this range. In this region, the density $\rho (\lambda )$ has an elevated
central part that differs noticeably from the semicircular distribution. The
spectrum also has a tiny tail given by Eq. (\ref{39b}) which can hardly be
seen in Fig. \ref{fig1}, see for detail Section V and Refs. \cite{rb88,sc02}


{\em Scale-free networks.} Spectra of scale-free graphs with the degree
distribution $P(k)=P_{0}k^{-\gamma }$ differ strongly from the semicircular
law \cite{fdbv01,gkk01}. The Barab\'{a}si-Albert\ model has a tree-like
structure, the exponent $\gamma =3$ of the degree distribution, and
negligibly weak correlations between degrees of the nearest neighbors \cite
{ab01a}. Therefore, one can assume that the spectrum of a random tree-like
graph can mimic well the spectrum of the model. In Fig. \ref{fig1} we
compare the spectrum of the random tree-like graph with $\gamma =3$ and the
spectrum of the Barab\'{a}si-Albert model obtained from simulations \cite
{fdbv01}. The density of states $\rho (\lambda )$ has a triangular-like form
and demonstrates a power-law tail. There is only a noticeable deviation of
the EM results from the results of simulations \cite{fdbv01} at small
eigenvalues $\lambda $. \ In order to improve the EM results, we used, as an
ansatz, the distribution function $F_{\lambda }(x)=[1+a(\lambda
)x^{2}]e^{-ixT(\lambda )}$ instead of the function $F_{\lambda
}(x)=e^{-ixT(\lambda )}$. In this case, there are two unknown complex
functions $a(\lambda )$ and $T(\lambda )$ which were determined
self-consistently from Eq. (\ref{34}).

{\em Power-law tail. }The power-law behavior of the density of eigenvalues $%
\rho (\lambda )\varpropto \lambda ^{-\delta }$ is an important feature of
the spectrum of scale-free networks. The simulations \cite{fdbv01} of the
Barab\'{a}si-Albert model having the degree exponent $\gamma =3$ revealed a
power-law tail of the spectrum, with the eigenvalue exponent $\delta
\approx5 $. Our prediction $\delta =2\gamma -1=5$ is in agreement with the
result of these simulations.

The study of the topology of the Internet at the Autonomous System (AS)
level revealed a power-law behavior of eigenvalues of the associated
adjacency matrix \cite{fff99,sfff03}. The degree distribution of the network
has the exponent $\gamma \approx 2.1$\cite{sfff03}. The eigenvalues $\lambda
_{i}$ of the Internet graph are proportional to the power of the rank $i$ of
an eigenvalues (starting with the largest eigenvalue): 
$\lambda _{i}\varpropto i^{\epsilon }$ with some exponent $\epsilon $. This
leads to $\rho (\lambda )\varpropto \sum_{i}\delta (\lambda -\lambda
_{i})\varpropto \lambda ^{-1+1/\epsilon }$. The {\it Multi} dataset analyzed
in \cite{sfff03} gave $\epsilon \approx -0.447$ and, hence, the eigenvalue
exponent $1-1/\epsilon \approx 3.2$. The {\it Oregon} dataset \cite{sfff03}
gave $\epsilon \approx -0.477$, $1-1/\epsilon \approx 3.1$.

Our results with $\gamma =2.1$ substituted, give the eigenvalue exponent $%
\delta =2\gamma -1\approx 3.2$ 
in agreement with the results obtained from empirical data for this network.
There are the following reasons for the agreement between the theory for
tree-like graphs and the data for the Internet. At first, although the
average clustering coefficient of the Internet at AS level is about 0.2, the
local clustering coefficient rapidly decreases with increasing the degree of
a vertex \cite{vpv02}. In other words, the closest neighborhood of vertices
with large numbers of connections is ``tree-like''. Recall that vertices
with large numbers of connections determine the large-eigenvalue asymptotics
of the spectrum. So, we believe that our results for the asymptotics of the
spectra of tree-like networks is also valid for the Internet and other
networks with similar structure of connections. Secondly, the Internet is
characterized by strong correlations between degrees of neighboring vertices 
\cite{pvv01}. However, as we have shown in the Section VI, such short-range
degree correlations do not affect the power-law behavior of eigenvalues.

The study of the Internet topology \cite{sfff03} also revealed a
correspondence between the large eigenvalues $\lambda _{i}$ and the degree $%
k_{i}$: $k_{i}=\lambda _{i}^{2}$. This result is in agreement with our
theoretical prediction that it is the highly connected vertices with a
degree about $k_{\lambda }\approx \lambda ^{2}$ \ that produce the power-law
tail $\rho (\lambda )\varpropto \lambda ^{-\delta }$.

The calculations of the eigenvalues spectrum of the adjacency matrix of a
pseudofractal graph with $\gamma =2.585\ldots $ \cite{dgm02} have revealed a
power-law behavior with $\delta \approx 4.6$. The effective medium
approximation gives lower value $\delta =2\gamma -1\approx 4.2$. The origin
of the difference is not clear. One should note that the pseudofractal is a
deterministically growing graph with a very large clustering coefficient $%
C=4/5$ and, what is especially important, with long-range correlations
between degrees of vertices.


\begin{figure}
\epsfxsize=69mm
\centerline{\epsffile{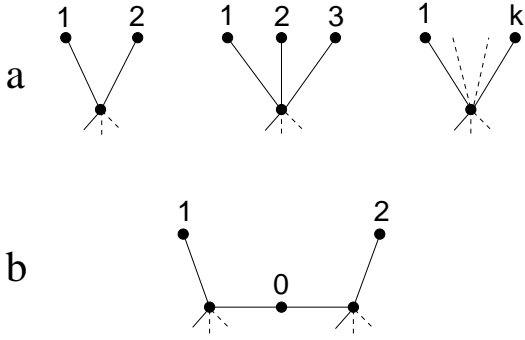}}
\caption{ Local configurations which produce localized states with
zero-eigenvalue (the central peak) in the spectrum of a random graph (see
the text). }
\label{fig3}
\end{figure}


{\em Weakly connected nodes.} Let us study the influence of weakly connected
vertices with degrees $1\leqslant k\leqslant 5$ on the spectra of random
tree-like graphs with the degree distribution $P(k)=P_{0}k^{-\gamma }$. In
Fig. \ref{fig2}$a$ and \ref{fig2}$b$ we represent the evolution of the
spectrum of the network with $\gamma =5$, when the smallest degree $k_{0}$
decreases from 5 to 1. The spectra were calculated in the framework of the
EM approximation. 
Similar results are obtained at different $\gamma $. For $k_{0}\leqslant 4$,
two peaks at non-zero eigenvalues emerge in the density of states $\rho
(\lambda )$. In order to understand an origin of the peaks one can note that
for this degree distribution the average degree $\left\langle k\right\rangle 
$ is close to $k_{0}$. For example, at $k_{0}=3$ we have $\left\langle
k\right\rangle =3.49$. Therefore, in this network, the probability to find a
vertex having three links is larger than the probability to find a vertex
with a degree $k\geqslant 4$. There are large parts of the network which
have a local $k=3$--regular structure. In Fig. \ref{fig2}$a$ we show a
density of eigenvalues of an infinite $k=3$--regular Bethe lattice [see Eq. (%
\ref{20}) at $k=3$]. At small eigenvalues, the density of the regular tree
fits well the density of the random network. At large $\lambda $, the
density of eigenvalues demonstrates a power-law behavior with the exponent $%
\delta =2\gamma -1$. 

In the case $k_{0}=2$ we have $\left\langle k\right\rangle =2.23$. This
network contains long chains which connect vertices with degrees $k\geqslant
3$. In Fig. \ref{fig2}$a$ we display the density of eigenvalues of an
infinite chain (see Eq. (\ref{20}) at $k=2$). At small eigenvalues this
density of eigenvalues fits well the density of eigenvalues of the random
network. Therefore, it is the vertices with small degrees that are
responsible for the formation of density $\rho (\lambda )$ of networks at
small eigenvalues.

{\em Dead-end vertices.} 
Let us investigate the effect of dead-end vertices on the spectra of random
tree-like graphs with different degree distributions. Fig. \ref{fig2}$b$
shows a spectrum of a scale-free network with $\gamma =5$ and the
probability of dead-end vertices $P(1)=0.3$. The EM approximation is used.
The spectrum has a flat part and two peaks at moderate eigenvalues. As we
have shown above, this (intermediate) part of the spectrum is formed mainly
by the vertices with degree $k=2$ and 3. 
The emergency of a dip at zero is a new feature of the spectrum. In fact,
there is a gap in the spectrum obtained in the in the framework of the EM
approach. The width of the gap increases with increasing $P(1)$. One can see
this in the insert on the Fig. \ref{fig2}$b$. The dead-end vertices also
produce a delta peak at $\lambda =0$. The central peak corresponds to
localized eigenstates.

Note that the appearance of the central peak and a dip is a general
phenomena in random networks with dead-end vertices. We also observed this
effect in the classical random graphs. Spectral analysis of the Internet
topology on the AS level revealed a central peak with a high multiplicity 
\cite{vhe02}. Thus the conjecture that localized and extended states are
separated in energy may well hold in complex networks. A similar spectra was
observed in many random systems, for example, in a binary alloy \cite{ke72}.
In order to estimate the height of the delta peak it is necessary to take
into account all localized states. Unfortunately, so far this is an unsolved
analytical problem \cite{vhe02}. In Fig. \ref{fig3} we show local parts of a
network, which produce localized states. One can prove that configurations
with two and more dead-end vertices, see Fig. \ref{fig3}$a$, produce
eigenstates with $\lambda =0$ . The corresponding eigenvectors have non-zero
components only at the dead-end vertices \cite{vhe02,hls03}. Fig. \ref{fig3}$%
b$ shows another configuration which produces an 
eigenstate with the eigenvalue $\lambda =0$. A corresponding eigenvector is
localized at vertices 0, 1 and 2.

{\em Finite-size effects.} In the present paper we studied the spectral
properties of infinite random tree-like graphs. Numerical studies of large
but finite random trees demonstrate that the spectrum of a finite tree
consists, speaking in general terms, of a continuous component and an
infinity of delta peaks. The components correspond to extended and localized
states, respectively \cite{g03}. There is a hole around each delta peak in
the spectrum. A finite regular tree has a spectral distribution function
which looks like a singular Cantor function \cite{hls03}. These results
demonstrate that finite size effects in spectra may be very strong. 
In particular, the finite size of a network determines the largest
eigenvalue in its spectrum. 
As was estimated in the Section V, the largest eigenvalue of the adjacency
matrix associated with a scale-free graph is of the order of $%
k_{cut}^{1/2}=k_{0}^{1/2}N^{1/2(\gamma -1)}$.

\begin{figure}
\epsfxsize=82mm
\centerline{\epsffile{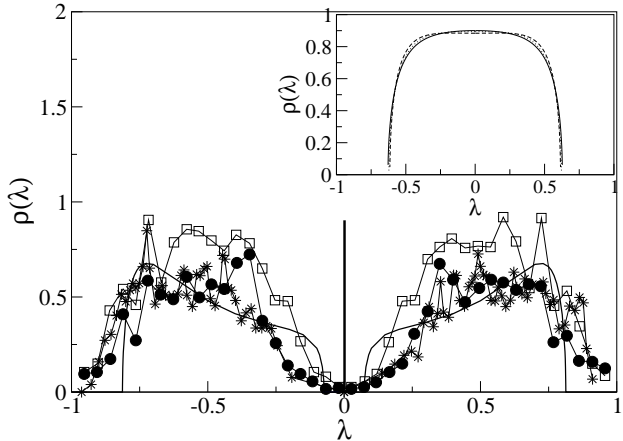}}
\caption{ Density of eigenvalues of the transition matrix $\widehat{P}$
defined by Eq. (\ref{61}). (i) The Internet data (the solid circles) and the
results of simulations of a random scale-free networks (the open squares)
from Ref. \protect\cite{esms02}. (ii) The spectrum of the Internet extracted
from Ref. \protect\cite{vhe02} (the stars). (iii) Our calculations (the
solid line) with the degree distribution $P(k)=Ak^{-2.1}$ for $k\geqslant 4$%
, $P(1)=0.358,$ \ \ $P(2)=0.4$ \ \ $P(3)=0.12$. These parameters are taken
from Ref. \protect\cite{vpv02}. The height of the central peak was estimated
from Ref. \protect\cite{vhe02}. The insert shows the spectra of the
transition matrix of a random tree-like graph with excluded dead-end
vertices: (i) a random tree-like graph with a scale-free degree
distribution, $\protect\gamma =3$ , the smallest degree $k_{0}=5$, and $%
\left\langle k\right\rangle =9.06$ (the dashed line). (ii) A classical
random graph with $\left\langle k\right\rangle =9.06$ (the dotted line).
(iii) A $k-$regular Bethe lattice with $k=9$ (the solid line). }
\label{fig4}
\end{figure}

{\em Spectrum of the transition matrix.} In Fig. \ref{fig4} we represent a
spectrum of the transition matrix $\widehat{P}$ defined by Eq. (\ref{61})
for a tree-like graph with the scale-free degree distribution $%
P(k)\varpropto k^{-\gamma }$ at large degrees $k\geqslant 5$. The spectrum
was calculated from Eqs. (\ref{66}) and (\ref{67}) with the degree exponent $%
\gamma =2.1$ and the probabilities $P(1),P(2),P(3)$ and $P(4)$ taken from 
empirical degree distribution of the Internet at the AS level \cite{vpv02}. 


The spectrum lies in 
the range $\left| \lambda \right| \leqslant \lambda _{2}<1$. In Fig. \ref
{fig4} we compare our results with the spectrum of the transition matrix $%
\widehat{P}$ of the Internet obtained in \cite{esms02,vhe02}. Unfortunately,
the data \cite{esms02,vhe02} are too scattered to make a detailed comparison
with our results. Nevertheless, one can see that the spectrum of $\widehat{P}
$ of the tree-like graph reproduces satisfactory the general peculiarities
of the real spectrum. Namely, the spectra have a wide dip at zero eigenvalue
and a central delta-peak \cite{vhe02}. The multiplicity of the zero
eigenvalue have been estimated in \cite{vhe02}. For a detailed comparison
between the spectra, correlations in the Internet must also be taken into
account.

In order to reveal an effect of dead-end vertices we calculated spectra of $%
\widehat{P}$ on a random tree-like graph with the Poisson and the scale-free
degree distributions $\gamma =3$ in the case when dead-end vertices are
excluded, that is $P(1)=0$, and $\left\langle k\right\rangle =9.06$. These
spectra are displayed in the insert on the Fig. \ref{fig4}. In the whole
range of eigenvalues these spectra are very close to the spectrum of a $k-$%
regular Bethe lattice with the degree $k=9.$ These calculations confirm the
fact that it is the dead-end vertices that produce the dip in the spectrum
of the Internet.

\section{Conclusions}

In this paper we have studied spectra of the adjacency and transition
matrices of random uncorrelated and correlated tree-like complex networks.
We have derived exact equations which describe the spectrum of random
tree-like graphs, and proposed a simple approximate solution in the
framework of the effective medium approach. Our study confirms that spectra
of scale-free networks as well as the spectra of classical random graphs do
not satisfy the Wigner law.

We have demonstrated that the appearance of a tail of the density of the
eigenvalues of sparse random matrices is a general phenomenon. The spectra
of classical random graphs (the Erd\H{o}s-R\'{e}nyi model) have a rapidly
decreasing tail. Scale-free networks demonstrate a power-law\ behavior of
the density of eigenvalues $\rho (\lambda )\propto \left| \lambda \right|
^{-\delta }$ . We have found a simple relationship between the degree
exponent $\gamma $ and the eigenvalue exponent $\delta $: $\delta =2\gamma
-1 $. We have shown that correlations between degrees of neighboring
vertices do not affect the power-law behavior of eigenvalues. Comparison
with the available results of the simulations of the Barab\'{a}si-Albert
model and the analysis of the Internet at the Autonomous System level shows
that this relationship is valid for these networks. We found that large
eigenvalues $\lambda \gg 1$ are produced by highly connected vertices with a
degree $k\approx \lambda ^{2}$.

Many real-world scale-free networks demonstrate short-range correlations
between vertices \cite{rsmob02,v03} and a decrease of a local clustering
coefficient with increasing degree of a vertex. Therefore, the relationship $%
\delta =2\gamma -1$ between the degree-distribution exponent $\gamma $ and
the eigenvalue exponent $\delta $ may also be valid for these networks. We
can conclude that the power-law behavior $\rho (\lambda )\varpropto \lambda
^{-\delta }$ is a general property of real scale-free networks.

Weakly connected vertices form the spectrum at small eigenvalues. Dead-end
vertices play a very special role. They produce localized eigenstates with $%
\lambda =0$ (the central peak). They also produce a dip in the spectrum
around the central peak. 
In conclusion, we believe that our general results for the spectra of
tree-like random graphs are also valid for many real-world networks with a
tree-like local structure and short-range degree correlations.

\acknowledgements S.N.D, A.N.S. and J.F.F.M. were partially supported by the
project POCTI/99/FIS/33141. A.G. acknowledges the support of the NATO
program OUTREACH.

{\em Note added}.---After we have finished our work we have learned about a
recent mathematical paper Ref. \cite{clv03}, where large eigenvalues of
spectra of complex random graphs were calculated. The statistical ensemble
of graphs, which was considered in that paper, essentially differs from that
of our paper and has a different cutoff of the degree distribution, but the
asymptotics of spectra agree in many cases\vspace{10pt}.

\noindent$^{\ast }$ Email address: sdorogov@fc.up.pt\newline
$^{\dagger }$ Email address: goltsev@mail.ioffe.ru \newline
$^{\ddagger }$ Email address: jfmendes@fis.ua.pt \newline
$^{\S}$ Email address: samaln@mail.ioffe.rssi.ru



\end{multicols} 


\begin{references}
\bibitem{note}  Note that in a finite Cayley tree, vertices are
statistically inequivalent. 
Indeed, there are vertices deep within the graph and boundary vertices.
\end{references}
\end{document}